\documentclass [12pt,a4paper]{article}
\usepackage{isridef}  

\def\epsPeps{_\epsilon)_\epsilon}     
\def\Me{{\text{\emph{M}}}}    
\def\Mt{{\text{M}}}           

\newcommand{\HU}{\text{H}}    

\numberwithin{equation}{section}

\begin{document}

\title{\bf\vspace{-2.5cm} The sequence of ideas in a re-discovery of
                          the Colombeau algebras}

\author{
         {\bf Andre Gsponer}\\
         {\it Independent Scientific Research Institute}\\ 
         {\it Oxford, OX4 4YS, England}
       }

\date{SEQUEN.10 ~~ \today}

\maketitle

\begin{abstract}

This is a gentle introduction to Colombeau nonlinear generalized functions, a generalization of the concept of distributions such that distributions can freely be multiplied.  It is intended to physicists and applied mathematicians who prefer a `step-by-step approach' to a `top-down indoctrination.'

  No particular prerequisite knowledge is necessary --- and in less than one hour you should know everything you need to know and were afraid to ask about Colombeau algebras and their applications in physics... 

  A selected bibliography is appended, giving examples of applications to partial differential and wave equations, electrodynamics, hydrodynamics, general relativity, and quantum field theory.

\end{abstract}

\noindent The goal of this tutorial is to lead the reader to rediscover by himself the key ideas which led Colombeau to define the proper generalization of the concept of distributions such that multiplication is always possible and meaningful.

  The emphasis is on concepts and methods, and the intent is to convince the reader that working with Colombeau nonlinear\footnote{The adjective `nonlinear' emphasizes that Colombeau generalized functions form an algebra.} generalized functions (in short, $\mathcal{G}$-functions), which can be differentiated and multiplied freely, is not more complicated than working with the familiar $\mathcal{C}^\infty$-functions.

  Since everything is self-contained and kept simple there are only few references in the text.  On the other hand, a selected bibliography with references to major publications on the subject is appended at the end.

  While Colombeau's seminal books \cite{COLOM1984-,COLOM1985-} are still highly valuable, the most recent comprehensive textbook is \cite{GROSS2001-}.  Short summaries of the main features of the Colombeau theory are included in most publications cited in the bibliography.  Furthermore, an alternative primer on Colombeau algebras is given in \cite{GSPON2006B}.

\section{Regular and irregular distributions} 
\label{reg:0}

The discovery of the Colombeau algebras is certainly one of the great events of the Twentieth Century history of mathematics.  To understand how it came about, let us start from another great invention, that of the theory of distributions by Bochner, Sobolev, Mikusinski, and Schwartz.  Indeed, whereas a regular distribution is a functional $\phi(T)$ having the representation
\begin{equation} \label{seq:1}
              \phi(T) \DEF  \int dx~\phi(x) T(x),
                \qquad \forall T(x) \in \mathcal{D},
\end{equation}
where $\phi(x)$ is a locally integrable function\footnote{In simple words, a function is locally integrable if it is integrable on every compact set.} and $T(x)$ a `test' function,\footnote{$\mathcal{D}$ is the space $\mathcal{D}(\Omega)$ of $\mathcal{C}^\infty$ functions with compact support on an open subset $\Omega \subset \mathbb{R}$.  For simplicity of notation we write $\mathcal{C}$, $\mathcal{C}^m$, and $\mathcal{C}^\infty$ for the continuous, $m$-times continuously differentiable, and respectively smooth functions with compact support on $\Omega$.  We similarly write $\mathcal{C}_p$ for the piece-wise continuous functions.  Then $\mathcal{C}^\infty \subset  \mathcal{C}^m \subset\mathcal{C}^0 = \mathcal{C} \subset\mathcal{C}_p$.  We also tacitly assume that all integrations are over $\mathbb{R}$, and that all functions are extended to $\mathbb{R}$ by setting them equal to zero outside of $\Omega$.  Finally, we set $\mathbb{N}_0 = \{0,\mathbb{N}\}$.} there is no such representation for the Dirac `function' $\delta(x)$ which is defined by the functional
\begin{equation} \label{seq:2}
   \delta(T) \DEF T(0),
                \qquad \forall T(x) \in \mathcal{D}.
\end{equation}
Thus, before the theory of such singular distributions was invented, the only thing that could be done was to write, \emph{symbolically},
\begin{equation} \label{seq:3}
           \delta(T) = \int dx~\delta(x) T(x),
                \qquad \forall T(x) \in \mathcal{D},
\end{equation}
and to refer to the definition \eqref{seq:2} for the interpretation of \eqref{seq:3}.

\section{The abstract and sequential views of distributions}

   Schwartz showed that $\delta(x)$ can be interpreted as an element of the space $\mathcal{D}'$ of continuous linear functionals on $\mathcal{D}$, and its derivatives defined as the derivatives of these functionals.  That is, if $\gamma \in \mathcal{D}'$ is any distribution, its derivatives in the `distributional sense' are such that, $\forall T \in \mathcal{D}$ and $\rmD^n = \partial^n/\partial x^n$,
\begin{equation} \label{seq:4}
  \int dx ~(\rmD^n \gamma \bigr )(x)~T(x)  = (-1)^n
  \int dx ~\gamma(x) ~(\rmD^n T \bigr )(x).     
\end{equation}
Alternatively, following Mikusinski, \eqref{seq:3} can be written as the weak limit of a sequence of $\mathcal{C}^\infty$ functions $\delta_\epsilon$, that is,
\begin{equation} \label{seq:5}
        \delta(T) \DEF \lim_{\epsilon \rightarrow 0} 
                 \int dx~\delta_\epsilon(x) T(x) = T(0),
                \qquad \forall T(x) \in \mathcal{D}.
\end{equation}
Indeed, if $\delta_\epsilon(x)$ is any family of functions
\begin{equation} \label{seq:6}
     \delta_\epsilon(x) = \rho_\epsilon(x) 
      \DEF \frac{1}{\epsilon} \rho\Bigl(\frac{x}{\epsilon}\Bigr),
\end{equation}
where $\epsilon \in ]0,1[$ is a parameter, and $\rho$ taken in the set\footnote{$\mathcal{S}$ is the space of $\mathcal{C}^\infty$ functions with steep descent, i.e., such that $f(x) \in \mathcal{S}$ and its derivatives decrease more rapidly than any power of $1/|x|$ as $x$ tends to infinity.  In distribution theory one usually takes $\rho \in \mathcal{D} \subset \mathcal{S}$ because this enable to deal with distributions with non-compact support unrestrictedly.}
\begin{equation} \label{seq:7}
\mathcal{A}_{0} \DEF \Bigl\{
    \rho(x) \in \mathcal{S},
    \quad \text{and} \quad
    \int dx~\rho(x) = 1 \Bigr\},
\end{equation}
making the change of variable $x=\epsilon y$ and taking the limit, it comes
\begin{equation} \label{seq:8}
    \int dx~\delta(x) T(x) \DEF \lim_{\epsilon \rightarrow 0} 
                 \int dy~\rho(y) T(\epsilon y) = T(0).
\end{equation}
Returning to equation \eqref{seq:6} one can observe that the sequence $\delta_\epsilon$ representing the $\delta$-distribution can actually by written as the convolution
\begin{equation} \label{seq:9}
    \delta_\epsilon(x) =  
                 \int dy~\rho_\epsilon(x-y) \delta(y) = \rho_\epsilon(x),
\end{equation}
provided the symbol $\delta$ inside the integral is interpreted according to its functional definition \eqref{seq:2}.

   This method is general:  It can be proved that convoluting a regular or singular distribution with any $\rho_\epsilon$ provides a \emph{representative sequence}
\begin{equation} \label{seq:10}
    \gamma_\epsilon(x) \DEF \rho_\epsilon(x) \ast \gamma(x) 
                          = \int dy~\rho_\epsilon(x-y) \gamma(y),
\end{equation}
of that distribution.  As $\gamma_\epsilon(x) \in \mathcal{C}^\infty$ this process of generating a smooth representative of $\gamma(x) \in \mathcal{D}'$ is called a \emph{regularization}, and the regularizing functions defined by \eqref{seq:7} are termed regularizers or \emph{mollifiers}.
  Thus, if $\gamma$ is any regular or singular distribution, \eqref{seq:1} can be written 
\begin{equation}
\label{seq:11}
        \gamma(T) \DEF \lim_{\epsilon \rightarrow 0} 
                 \int dx~\gamma_\epsilon(x) ~T(x),
                \qquad \forall T(x) \in \mathcal{D},
\end{equation}
and $\gamma$ is then interpreted as the equivalence class of the weakly convergent sequences of the smooth functions $\gamma_\epsilon$ modulo weak zero-sequencies.

   In comparison to Schwartz's abstract theory, the advantages of Mikusinski's sequential view are that it provides explicit representations for the distributions, and that their derivatives are simply obtained by differentiating the representative sequencies.

\section{Schwartz's local structure theorem}

   A particularly important contribution of Laurent Schwartz is the formulation of his local structure theorem stating that \emph{``any distribution is locally a partial derivative of a continuous function''} \cite[Theorems XXI and  XXVI]{SCHWA1966-}:
\begin{theorem}
 Let $\mathcal{D}'(\Omega)$ be the space of distributions on the compact set $\Omega$. Then every $\gamma \in \mathcal{D}'$ is of the form
\begin{equation} \label{seq:12}
   \gamma(x) = \sum_n \rmD^n g_n(x),     
\end{equation}
where  $n \in \mathbb{N}_0$, and the support of each $g_n  \in \mathcal{C}(\Omega)$ is contained in an arbitrary compact neighborhood $K \subset \Omega$.
\end{theorem}
For example, Dirac's function $\delta(x)$ is generated by the second distributional derivative of the absolute value $|x| \in \mathcal{C}^0$, i.e.,  $\delta(x) = 1/2~\rmD^2 |x|$.\\

   Differentiation induces therefore the following remarkable cascade of relationships: {\it continuously differentiable functions $\rightarrow$ continuous functions $\rightarrow$  distributions}. This gives a unique position to Schwartz distributions because they constitute the smallest space in which all continuous functions can be differentiated any number of times.  For this reason it is best to reserve the term `distribution' to them, and to use the expression `generalized function' for any of their generalizations.  On the other hand, classical generalizations of the concept of function such as piece-wise continuous functions, measures, Cauchy and Hadamard finite-parts of integrals, etc., are all distributions.

    For application in physics Schwartz's structure theorem is of great significance because it asserts that singular distributions do not come `from nowhere,' but derive from a generating function $g(x) \in \mathcal{C}^0$.  For example, the classical electron charge distribution originates from the absolute value in the definition of the Coulomb potential, i.e., $\phi = e/|\vec{r}\,|$, and due consideration to this fact leads to a distributionally consistent introduction of point charges and dipoles in classical electrodynamics \cite{TANGH1962-,GSPON2006B2,GSPON2004D,GSPON2007C,GSPON2008B}.

\section{Schwartz's multiplication impossibility theorem}

Distributions generalize ordinary functions, which can be regarded as trivial cases of distributions.  They enjoy most of the properties of $\mathcal{C}^\infty$ functions (e.g., they can be differentiated any number of time) with the notable exception of \emph{multiplication}.

  For example, if the product of distributions is defined in the most natural way, i.e., by multiplying representative sequencies, the square of the $\delta$-function corresponds to $(\delta^2)_\epsilon(x) = (\delta_\epsilon)^2(x) = \rho_\epsilon^2(x)$.  Then, when evaluated on a test function $T\in\mathcal{D}$ according to \eqref{seq:11}, we get,
\begin{equation} \label{seq:13}
    \int dx~\delta^2(x) T(x) \DEF \lim_{\epsilon \rightarrow 0} 
                 \int dx~\rho_\epsilon^2(x) T(x)
   = \lim_{\epsilon \rightarrow 0} \frac{T(0)}{\epsilon}
                 \int dy~\rho^2(y)  = \infty,
\end{equation}
which implies that $\delta^2$ is \emph{not} a distribution.  Many mathematicians have of course tried to define a consistent product of distributions.  But these efforts only confirmed that there is no multiplication on all of $\mathcal{D}'$ which still has values in $\mathcal{D}'$, unless some  essential properties are given up.  For instance, in any formulation such that the usual relations $x \cdot 1/x =1 $ and $x \cdot \delta(x)=0$ hold, associativity leads to the contradiction
\begin{equation} \label{seq:14}
    \Bigl( x \cdot \frac{1}{x} \Bigr) \cdot \delta(x) = \delta(x),
   \qquad \text{whereas} \qquad 
     \frac{1}{x} \cdot \Bigl(x \cdot \delta(x) \Bigr) = 0.
\end{equation}

   The goal therefore shifted towards finding an algebra $\mathcal{G}$ of generalized functions containing the distributions and preserving most of the desirable properties of ordinary functions.  But even that less ambitious goal turned out to be quite difficult.  In particular, there are many options and it is not possible to know a priori which `essential properties' should be preserved.  For instance, possibly inspired by the cardinal position of continuous functions in his structure theorem \eqref{seq:12}, Laurent Schwartz was particularly attached to the idea that these functions should have a similar position in $\mathcal{G}$.  He therefore postulated a set of minimum requirements which can be phrased as follows\footnote{See for example  \cite[p.6]{GROSS2001-}.}
\begin{enumerate}

\item[{\fbox{0}}] The differential algebra $\mathcal{G}$ is associative and commutative.  Its elements are written $[u]$ when it is useful to emphasize that $u \in \mathcal{G}$.
 
\item[{\fbox{1}}] The space of distributions $\mathcal{D}'$ is linearly embedded into $\mathcal{G}$, and the function $f(x)\equiv 1$ is the unit element for their product `$\odot$' in $\mathcal{G}$, i.e.,  $\forall \gamma \in \mathcal{D}'$, there is an embedding $\mathcal{D}' \rightarrow \mathcal{G}, \gamma \mapsto [\gamma]$, and ~$[1] \odot [\gamma] = [\gamma]$;

\item[{\fbox{2}}] There exists a derivation operator $\rmD: \mathcal{G} \rightarrow \mathcal{G}$ that is linear and satisfies the Leibniz rule, i.e., $\forall u,v \in \mathcal{G}, ~\rmD(u\odot v) = (\rmD u)\odot v + u\odot(\rmD v)$;

\item[{\fbox{3}}] $\rmD$ restricted to $[\mathcal{D}']$ is the usual partial derivative consistent with the integration by parts formula \eqref{seq:4};

\item[{\fbox{4}}] The product of two continuous functions embedded in $\mathcal{G}$ coincides with the usual pointwise product `$\cdot$' in $\mathcal{C}$, i.e., $\forall f,g \in \mathcal{C}, ~[f] \odot [g] = f\cdot g$.
\end{enumerate}
Unfortunately, on the basis of simple counter-examples, it is easy to show that \emph{there is no associative and commutative differential algebra $\mathcal{G}$ satisfying the requirements} {\fbox{1}} -- {\fbox{4}}.  For example, they lead to the conclusion $\rmD^2 |x| = 0$, whereas, as recalled above, $\rmD^2 |x| = 2 \delta(x)$ in distribution theory.  This is the famous Schwartz impossibility theorem of 1954.

\section{Colombeau's breakthrough}

  It was only in 1983 that Jean-Fran\c{c}ois Colombeau was able to show in a truly satisfactory manner that it is actually possible to construct associative and commutative algebras satisfying {\fbox{1}} -- {\fbox{3}}, provided {\fbox{4}} is replaced by:\footnote{The original discoveries of Colombeau were made in a different context, and arose from more abstract considerations. But their success can be traced to the emphasis given to $\mathcal{C}^\infty$ rather than to continuous functions in general, an emphasis which may have a deep physical significance.}
\begin{enumerate}
\item[{\fbox{4'}}] The product of two $\mathcal{C}^\infty$ functions embedded in $\mathcal{G}$ coincides with the usual pointwise product `$\cdot$' in $\mathcal{C}^\infty$, i.e., $\forall f,g \in \mathcal{C}^\infty, ~[f] \odot [g] = f\cdot g$.
\end{enumerate}
Therefore, since $\mathcal{C}^\infty \subset \mathcal{C}$, it was by relaxing the requirement {\fbox{4}} that it became possible to move forward:  As $\mathcal{C}^\infty$ functions have much more powerful properties than continuous functions in general, e.g., Taylor's theorem with remainder, the problem became manageable.

\section{The embedding space}

   Colombeau's axiom {\fbox{4'}} combined with axioms {\fbox{1}} -- {\fbox{3}} implies that  $\mathcal{G}$ contains $\mathcal{C}^\infty$ as a differential subalgebra:  This opens the way to the possibility that $\mathcal{G}$ could be similarly contained in a larger differential algebra $\mathcal{E}$ such that its elements would be $\mathcal{C}^\infty$ in the variable $x$.  Since the mollified sequencies $\gamma_{\epsilon}(x)$ representing the distributions are precisely $\mathcal{C}^\infty$ in the variable $x$, this suggests to define $\mathcal{E}$, the \emph{embedding space}, as the set of maps,\footnote{The notation  $(f\epsPeps$ where $0 <\epsilon < 1$, which will be later abbreviated as $f_\epsilon$, emphasizes that $(f\epsPeps$ is an element of $\mathcal{E}$ rather than a usual representative sequence \eqref{seq:10}.} 
\begin{align}
\nonumber
    \mathcal{E} \DEF  \Bigl\{
     (f\epsPeps :
     \mathcal{A}_q \times \Omega &\rightarrow \mathbb{R},\\
\label{seq:15}
                             (\eta, x) &\mapsto 
                                        (f\epsPeps(\eta,x)
                      \Bigr\},
\end{align}
which are $\mathcal{C}^\infty$ functions in the variable $x\in\Omega$ for any given \emph{Colombeau mollifier} $\eta \in \mathcal{A}_q$, where $\mathcal{A}_q \subset \mathcal{A}_0$ remains to be specified, and which depend on the parameter $\epsilon \in ]0,1[$ through the scaled mollifier  
\begin{equation} \label{seq:16}
     \eta_\epsilon(x) 
      \DEF \frac{1}{\epsilon} \eta\Bigl(\frac{x}{\epsilon}\Bigr).
\end{equation}
Obviously, $\mathcal{E}$ is an associative and commutative differential algebra with unit $(\eta, x) \mapsto 1$ with respect to pointwise multiplication. It contains $\mathcal{C}^\infty$ as the subset of the maps \eqref{seq:15} which do not depend on $\eta$, i.e., $(f\epsPeps(x) \equiv f(x)$.

  The distributions $f \in \mathcal{D}'$ are then embedded in $\mathcal{E}$ as the convolutions\footnote{This definition due to Colombeau differs by a sign from the usual definition \eqref{seq:10} of regularization.} 
\begin{align}
\nonumber
          (f\epsPeps(x)  \DEF \eta_\epsilon(-x) \ast f(x)
                        &=    \int \frac{dy}{\epsilon}
                              \eta\Bigl( \frac{y-x}{\epsilon} \Bigr) f(y)\\
\label{seq:17}   
                        &=    \int dz~\eta(z) ~f(x + \epsilon z),
\end{align}
where, in order to define $\mathcal{G} \subset \mathcal{E}$, the Colombeau mollifiers $\eta \in \mathcal{A}_q$ may need to have specific properties in addition to those implied by \eqref{seq:7}.  In particular, since $\mathcal{C}^\infty \subset \mathcal{D}'$ there are two distinct embeddings of $\mathcal{C}^\infty$ in $\mathcal{E}$:  Its embedding by \eqref{seq:17} as a subset of $\mathcal{D}'$, and its direct inclusion by \eqref{seq:15} according to the maps $(f\epsPeps(x) \equiv f(x)$.  To be consistent with axiom {\fbox{4'}}, the mollifiers $\eta \in \mathcal{A}_q$ have thus to be such that $[(f\epsPeps](x) = [f](x) = f(x)$ for all $f \in \mathcal{C}^\infty$.

\section{Embedding of $\mathcal{C}^\infty$ functions}

   To find these additional properties we begin by studying the embeddings and products of $\mathcal{C}^\infty$ functions.  We therefore calculate \eqref{seq:17} for $f \in \mathcal{C}^\infty$ and apply Taylor's theorem to obtain at once
\begin{align}
 \label{seq:18}
    (f\epsPeps(x) 
  &= f(x) \int dz~\eta(z) + ...\\ 
 \label{seq:19}
  &+ \frac{\epsilon^n}{n!} f^{(n)}(x) \int dz~z^n \eta(z)+ ...\\
 \label{seq:20}
  &+ \frac{\epsilon^{(q+1)}}{(q+1)!}  \int dz~z^{q+1}\eta(z)
                           ~f^{(q+1)}(x+ \vartheta\epsilon z),
\end{align}
where $f^{(n)}(x)$ is the $n$-th derivative of $f(x)$, and $\vartheta \in ]0,1[$.  Then, since $\eta \in \mathcal{S}$ and $f$ has a compact support, the integral in \eqref{seq:20} is bounded so that the remainder is of order $\OOO_x(\epsilon^{q+1})$ at any fixed point $x$.

  Moreover, if following Colombeau the mollifier $\eta$ is chosen in the set
\begin{equation} \label{seq:21}
\mathcal{A}_{q} \DEF \Bigl\{
    \eta(x) \in \mathcal{A}_{0},
    \quad \text{and} \quad
    \int dz~z^n\eta(z) = 0, \quad \forall n=1,...,q \Bigr\},
\end{equation}
all the terms in \eqref{seq:19} with $n \in [1,q]$ are zero and we are left with
\begin{align}
 \label{seq:22}
   \forall f \in \mathcal{C}^\infty, \qquad
    (f\epsPeps(x) = f(x) + \OOO_x(\epsilon^{q+1}).
\end{align}
Therefore, provided the set $\mathcal{A}_q$ is not empty and $q$ can take any value in $\mathbb{N}$, it is possible to make the difference $(f\epsPeps(x) - f(x)$ as small as we please even if $\epsilon \in ]0,1[$ is kept \emph{finite}.   If we now consider a product of two $\mathcal{C}^\infty$ functions, it is easily seen that equation \eqref{seq:22} immediately leads to
\begin{align}
 \label{seq:23}
   \forall u,v \in \mathcal{C}^\infty, \qquad
    (u\epsPeps(x) \cdot (v\epsPeps(x)
   = u(x) \cdot v(x) + \OOO_x(\epsilon^{q+1}),
\end{align}
where the remainder $\OOO_x(\epsilon^{q+1})$ is still as small as we please for any $\epsilon \in ]0,1[$ if $q$ is large enough.

\section{Colombeau mollifiers and Fourier transformation}

Colombeau proved that $\mathcal{A}_q$ is not empty and provided a recursive algorithm for constructing the corresponding mollifiers for all $q \in \mathbb{N}$.  He also showed  \cite[p.7]{COLOM1985-}, \cite[p.113]{COLOM1985-}, \cite[p.169]{COLOM1992-} that due to the Fourier transformation identities
\begin{align}
 \label{seq:24}
    \int dx~\eta(x) = \FOU{\eta}(0),
    \qquad \text{and} \qquad
    \int dx~x^n\eta(x) = (-i)^n \frac{d^n \FOU{\eta}}{dp^n}(0), 
\end{align}
%
%
which are valid $\forall \eta \in \mathcal{S}$, the conditions \eqref{seq:21} on the moments of $\eta(x)$ can be replaced by equivalent conditions on the derivatives of its Fourier transform $\FOU{\eta}(p)$.  Thus by taking for $\eta(x)$ any real functions such that $\FOU{\eta}(p)\equiv 1$ in a finite neighborhood of $p=0$, one automatically satisfies the conditions \eqref{seq:21} for \emph{any} $n \in \mathbb{N}$, that is for $q$ as large as we please.  For this reason the set of mollifiers
%
\begin{equation} \label{seq:25}
\mathcal{A}_\infty \DEF \Bigl\{
    \eta(x) \in \mathcal{S}, 
    \quad \text{such that} \quad
    \FOU{\eta}(0)\equiv 1 \Bigr\}.
\end{equation}
is written $\mathcal{A}_\infty$.  In this paper all Colombeau mollifiers will be taken in that set.

For example, with $\eta \in \mathcal{A}_\infty$ the Colombeau embeddings of any two polynomials, and the products of these embeddings, are identical to these polynomials and to their ordinary products.  That is, axiom {\fbox{4'}} is identically satisfied for polynomials.  But for the other $\mathcal{C}^\infty$ functions there will still be a remainder to be taken care of, even if it is infinitesimal.

\section{Embedding of continuous functions}

Let us now consider the embedding of continuous functions assuming that the only things that are known is that they are continuous and compactly supported.  Then, a priori, there is little more that can be done than writing
\begin{equation} \label{seq:26}
   \forall f \in \mathcal{C}, \qquad
   (f\epsPeps(x)
              =  \int dz~\eta(z) ~f(x + \epsilon z),
\end{equation}
because neither Taylor's formula nor the mean-value theorem can be applied to transform the right-hand side into a more useful expression.
In fact, the only fully general expression comparable to \eqref{seq:22} is 
\begin{equation} \label{seq:27}
   \forall f \in \mathcal{C}, \qquad
   (f\epsPeps(x)
              =  f(x) + \ooo_{x,\epsilon}(1),
\end{equation}
which simply means that $(f\epsPeps(x)$ converges uniformly to $f(x)$ as $\epsilon \rightarrow 0$ because $f$ has compact support.  Any more precise statement requires that the continuous function is further specified.

   For example, if $f$ is $m$-times continuously differentiable we can write
\begin{equation} \label{seq:28}
   \forall f \in \mathcal{C}^m, \qquad
   (f\epsPeps(x)
              =  f(x) + \OOO_x(\epsilon^m),
\end{equation}
where, in contrast to \eqref{seq:22}, $m \geq 1$ is a fixed integer.

\section{Embedding of distributions}

  We now turn to distributions.  By Schwartz's local structure theorem \eqref{seq:12} they can be written $\gamma(x) = \rmD^n g(x)$ where $g \in\mathcal{C}(K)$ if we restrict ourselves to a single generating function with support in a compact set $K$.  Then for $x\in K$ their embeddings \eqref{seq:17} are, using the integration by parts formula \eqref{seq:4},
\begin{align}
 \label{seq:29}
    (\gamma\epsPeps(x)   
   &= \int dz~ \eta(z) ~\rmD^n_x g(x + \epsilon z),\\
 \label{seq:30}
   & = \frac{1}{\epsilon^n}\int dz~\eta(z)
            ~\rmD^n_z g(x + \epsilon z),\\
 \label{seq:31}
   & = \Bigl(\frac{-1}{\epsilon}\Bigr)^n
       \int dz~\bigl(\rmD^n\eta\bigr)(z) ~g(x + \epsilon z).
\end{align}
Since $\eta \in \mathcal{S}$, and $g \in \mathcal{C}$ is compactly supported, the last integral is bounded
and we get
\begin{align}
 \label{seq:32}
   \forall \gamma \in \mathcal{D}', \qquad
    (\gamma\epsPeps(x) 
     = \OOO_x({1}/{\epsilon^{n}}),
   \qquad \text{as} \quad \epsilon \rightarrow 0.
\end{align}
This bound is compatible with the bounds (\ref{seq:22}, \ref{seq:27}, and \ref{seq:28}) because $\mathcal{D}'$ contains all continuous functions.  To illustrate its significance for non-trivial distributions we need to consider generating functions $g \in \mathcal{C}^0$.

   For example, we know that $\delta(x) = \rmD^2 g(x)$ with $g(x) = |x|/2$.  On the other hand, the Colombeau embedding \eqref{seq:17} of $\delta(x)$ is
\begin{align}
 \label{seq:33}
   (\delta\epsPeps(x) = \frac{1}{\epsilon}
                                 ~\eta\Bigl(-\frac{x}{\epsilon}\Bigr)
                                   = \OOO_x(1/\epsilon),
\end{align}
which has an $\epsilon$-dependent bound consistent with the bound \eqref{seq:32}, i.e., $1/\epsilon < 1/\epsilon^2$, although their exponents disagree by one unit.  This is because \eqref{seq:32} is fully general and thus does not take the particular properties of $g(x)$ into account.  In the present case it is easy to calculate $(|x|\epsPeps(x)$ with \eqref{seq:17} and to verify that $2(\delta\epsPeps(x) = \rmD^2( |x|\epsPeps(x) = \OOO_x(1/\epsilon)$ rather than $\OOO_x(1/\epsilon^2)$.  For the same reason the embedding of the Heaviside function $\HU(x) = \rmD |x|/2$ is
\begin{align}
 \label{seq:34}
   (\HU\epsPeps(x) = \int_{-\infty}^{x/\epsilon} dz~\eta(-z)
                                   = \OOO_x(1).
\end{align}
which, rather than $\OOO_x(1/\epsilon)$, has the $\epsilon$-dependence $\OOO_x(1)$ characteristic of a piece-wise continuous function because of its jump at $x=0$.

  To give another example, the singular distributions generated by the derivatives of the $\mathcal{C}^0$ function equal to $0$ for $x \leq 0$ and to $x^r$ for $x>0$, with $r \in ]0,1[$, have an $\epsilon$-dependent bound $\OOO_x(\epsilon^{r-n})$ where $r-n < 0$ in $\mathbb{R}$.

    In summary, \eqref{seq:32} provides a conservative bound for the $\epsilon$-dependence of all Schwartz distributions.  In the case of singular distributions, the exponent in the bound \eqref{seq:32} can be any integer $n \in \mathbb{N}$.

\begin{table}
\begin{center}
\begin{tabular}{|c|@{}c@{}|l|l|}
\hline
\multicolumn{4}{|c|}{\raisebox{+0.4em}{{\bf Local structure of embeddings and of their differences in $\mathcal{E}$ \rule{0mm}{6mm}}}} \\
\hline       
$f$                  &\,& ~~~ ~~ $(f\epsPeps(x)$ &    \rule{0mm}{5mm}\\
\hline
\hline        
$\mathcal{C}^\infty$ &\,& $f(x)$ & Directly included smooth function \rule{0mm}{5mm}\\
\hline
\hline
$\mathcal{C}^\infty$ &\,& $f(x) + \OOO_x(\epsilon^q)$ & Smooth function, $q > \forall p \in \mathbb{N}$  \rule{0mm}{5mm}\\
\hline
$\mathcal{C}$        &\,& $f(x) + \ooo_{x,\epsilon}(1)$ & Continuous function  \rule{0mm}{5mm}\\
\hline
$\mathcal{C}_p$      &\,& $~~~ ~~~ ~~~ ~~~\, \OOO_x(1)$ & Piece-wise continuous function  \rule{0mm}{5mm}\\
\hline
$\mathcal{D}'$       &\,& $~~~ ~~~ ~~~ ~~~\, \OOO_x(\epsilon^{-N})$ & Singular distribution, $ N \in \mathbb{N}$  \rule{0mm}{5mm}\\
\hline
\hline
$\mathcal{N}$        &\,& $~~~ ~~~ ~~~ ~~~\, \OOO_x(\epsilon^q)$ & Negligible function, $q > \forall p \in \mathbb{N}$  \rule{0mm}{5mm}\\
\hline
\end{tabular}

\caption{\emph{
The differential algebra $\mathcal{E}$ contains the smooth functions as direct embeddings $f \in \mathcal{C}^\infty \subset \mathcal{E}$, and also as mollified embeddings $(f\epsPeps \in (\mathcal{C}^\infty\epsPeps \subset \mathcal{E}$.
The embeddings $(f\epsPeps$ of the continuous functions and of the distributions are sorted in terms of the behavior of the bound on their $\epsilon$-dependent part as $\epsilon \rightarrow 0$. $f(x)$ is the point-value of the continuous functions at $\epsilon=0$.  The negligible functions are infinitesimally small elements such as the differences between the direct inclusions and the Colombeau-mollified embeddings of the $\mathcal{C}^\infty$ functions.
}} 
\label{table:1}
\end{center}
\end{table}

\section{Linear operations and negligible functions}

   In Table~\ref{table:1} the usual functions and the distributions are classified according to the structure of their embeddings in $\mathcal{E}$.  Referring to this table it is easy to predict the structure of the result of binary algebraic operations in $\mathcal{E}$, and thus to get clues on how to define the algebra $\mathcal{G}$. 

   For instance, in the last line of Table~\ref{table:1} the set denoted by $\mathcal{N}$ consists of functions which are not the direct result of embeddings:  It is the algebra of the so-called \emph{negligible functions},\footnote{A proper definition of negligible functions will be given shortly.} which arise in particular from subtracting the two different inclusions of the  $\mathcal{C}^\infty$ functions, i.e., 
\begin{equation} \label{seq:35}
   \forall f \in \mathcal{C}^\infty,
   \quad \forall q \in \mathbb{N},
   \qquad (f\epsPeps(x) -  f(x) = \OOO_x(\epsilon^q) \quad \in \mathcal{N}.
\end{equation}
Then for all linear operations (i.e., addition/subtraction, multiplication by a scalar, and differentiation) it is clear that the results will always be in one of the sets listed in the first column of Table~\ref{table:1}, which is therefore a suitable classification of the embeddings of the usual functions and distributions with regards to linear operations in $\mathcal{E}$.

   Of course, the negligible functions of the type \eqref{seq:35} are precisely the differences that are to be taken care of in order to satisfy axiom \fbox{4'}.  In particular, they will remain `negligible' as long as they are not multiplied by `very large' functions.  This is why we have now to look at the nonlinear operations  in $\mathcal{E}$.

\section{Nonlinear operations and moderate functions}

   We know that the product of two distributions (or of a continuous function and a distribution) will, in general, not be a distribution.  For example, the $n$-th power of the Dirac $\delta$-function can be defined by the $n$-th power of the embedding \eqref{seq:33}, i.e.,
\begin{align}
 \label{seq:36}
   (\delta^n\epsPeps(x) = \frac{1}{\epsilon^n}
             \eta^n\Bigl(-\frac{x}{\epsilon}\Bigr) = \OOO_x(\epsilon^{-n}).
\end{align}
But, despite that $(\delta^n\epsPeps(x)$ has a $\OOO_x(\epsilon^{-n})$ dependence similar to that of a `distribution,' it is not a distribution in the sense of Schwartz and Mikusinski --- rather, it is an element of a larger set of `generalized functions' containing the distributions as a subspace.

  This led Colombeau to define the set $\mathcal{E}_\Mt$, which he called \emph{moderate functions},\footnote{A proper definition of moderate functions will be given shortly.}
\begin{equation} \label{seq:37}
   \forall (g\epsPeps \in \mathcal{E}_\Mt~:
   \qquad \exists N \in \mathbb{N}_0,
      \qquad \text{such that} \qquad 
   (g\epsPeps(x) = \OOO_x(\epsilon^{-N}).
\end{equation}
It is evident that $\mathcal{N} \subset (\mathcal{C}^\infty)_\epsilon \subset (\mathcal{C})_\epsilon \subset (\mathcal{D}')_\epsilon \subset \mathcal{E}_\Mt$, and a matter of elementary calculations to verify that $\mathcal{E}_\Mt$ and $\mathcal{N}$ are algebras for the usual pointwise operations in $\mathcal{E}$, and that $\mathcal{N}$ is an ideal of $\mathcal{E}_\Mt$.   Indeed, the product of two moderate functions is still moderate --- they are \emph{multipliable} --- and as $q$ in \eqref{seq:35} is as large as we please, and $N$ in \eqref{seq:37} a fixed integer, the product of a negligible function by a moderate one will always be a negligible function.  Moreover, $\mathcal{E}_\Mt$ is a differential algebra satisfying axioms {\fbox{1}} -- {\fbox{3}}, and it is not difficult to show that $\mathcal{E}_\Mt$ is the largest differential subalgebra (i.e., stable under partial differentiation) of $\mathcal{E}$ in which $\mathcal{N}$ is a differential ideal.

   Furthermore, one can also consider infinite sums of products of moderate functions and take their limits in $\mathcal{E}$.  It is then easily verified that, for example, $\sqrt{(\delta^n\epsPeps}(x)$ and $\sin(\delta^n\epsPeps(x)$ are elements of $\mathcal{E}_\Mt$.  On the other hand
\begin{align}
 \label{seq:38}
   \exp(|\delta|\epsPeps(0)
 = \OOO_x\bigl(\exp(1/\epsilon)\bigr) \not\in \mathcal{E}_\Mt,
\end{align}
so that $\exp(\delta\epsPeps(x)$ is a \emph{non-moderate function}, and thus an element of the complement of $\mathcal{E}_\Mt$ in $\mathcal{E}$.  Conversely, $\exp(-|\delta|\epsPeps(x)$ is a negligible function, so that $\mathcal{N}$ contains elements of exponentially fast decrease.  

   Consequently, when operating in full generality in $\mathcal{E}$, that is when including multiplication and limiting processes,  one is led to consider its elements as in Table~\ref{table:2}, i.e., as members of the differential algebras $\mathcal{C}^\infty$, $\mathcal{N}$, and $\mathcal{E}_\Mt$, rather than as  members of the embeddings of the classical spaces $\mathcal{C}^\infty, \mathcal{C}$, and $\mathcal{D}'$ as in Table~\ref{table:1}.

\def\Mcompl{\mathcal{E}_\Mt\backslash\mathcal{E}}
\begin{table}
\begin{center}
\begin{tabular}{|c|@{}c@{}|c|c|c|c|}
\hline
\multicolumn{6}{|c|}{\raisebox{+0.4em}{{\bf Multiplication in $\mathcal{E}$\rule{0mm}{6mm}}}} \\
\hline        
{\bf ~}           & \, & ~~$\mathcal{C}^\infty$~~ & ~~$\mathcal{N}$~~  & ~~$\mathcal{E}_\Mt$~~ & 
                                                                   $\Mcompl$   \rule{0mm}{5mm}\\
\hline\hline
$\mathcal{C}^\infty$            & & $\mathcal{C}^\infty$    &  $\mathcal{N}$         & $\mathcal{E}_\Mt$ & 
                                                                    $\Mcompl$   \rule{0mm}{5mm}\\
\hline
$\mathcal{N}$        & & $\mathcal{N}$ & $\mathcal{N}$         & $\mathcal{N}$         & $\mathcal{E}$ \rule{0mm}{5mm}\\
\hline
$\mathcal{E}_\Mt$& & $\mathcal{E}_\Mt$ & $\mathcal{N}$ & $\mathcal{E}_\Mt$ & 
                                                                 $\mathcal{E}$   \rule{0mm}{5mm}\\
\hline
$\Mcompl$
                     & & $\Mcompl$     &  $\mathcal{E}$        & $\mathcal{E}$ & $\Mcompl$   \rule{0mm}{5mm}\\
\hline
\end{tabular}

\caption{\emph{
The elements of $\mathcal{E}$ remain in their respectives subalgebras $\mathcal{N}$, $\mathcal{E}_\Mt$, or $\Mcompl$ when multiplied by directly included $\mathcal{C}^\infty$ functions.  The subalgebra $\mathcal{N} \subset \mathcal{E}_\Mt$ is an ideal of $\mathcal{E}_\Mt$.  The products of negligible and moderate elements with elements in the complement of $\mathcal{E}_\Mt$ in $\mathcal{E}$ are in general undefined.
        }    }  
\label{table:2}
\end{center}
\end{table}

\section{Discovery of the Colombeau algebra}

The fact that $\mathcal{N}$ is an ideal of $\mathcal{E}_\Mt$ is the key to defining an algebra containing the distributions and satisfying axiom {\fbox{4'}}. Indeed, if we conventionally write $\mathcal{N}$ for any negligible function, then\footnote{From now on we abbreviate $(f\epsPeps$ as $f_\epsilon$.}
\begin{align}
 \label{seq:39}
   \forall g_\epsilon,h_\epsilon \in \mathcal{E}_\Mt, \qquad
   (g_\epsilon + \mathcal{N}) \cdot (h_\epsilon + \mathcal{N})
  = g_\epsilon \cdot h_\epsilon + \mathcal{N}.
\end{align}
Similarly, using the same convention, Eq.~\eqref{seq:23} giving the product of the Colombeau embeddings of two $\mathcal{C}^\infty$ functions can be written
\begin{align}
 \label{seq:40}
   \forall u,v \in \mathcal{C}^\infty, \qquad
   (u + \mathcal{N}) \cdot (v + \mathcal{N})
  = u \cdot v + \mathcal{N},
\end{align}
whereas axiom {\fbox{4'}} demands
\begin{align}
 \label{seq:41}
   \forall u,v \in \mathcal{C}^\infty, \qquad
   [u + \mathcal{N}] \odot [v + \mathcal{N}]
  = [u \cdot v + \mathcal{N}] = u \cdot v.
\end{align}
Thus, it suffice to define the elements of $\mathcal{G}$ as the elements of $\mathcal{E}_\Mt$ modulo $\mathcal{N}$, e.g., to identify $[g_\epsilon + \mathcal{N}]$ and $[g_\epsilon]$, so that $[u + \mathcal{N}]=[u]$ because $(\mathcal{C}^\infty)_\epsilon \subset \mathcal{E}_\Mt$, and axiom {\fbox{4'}} will be satisfied.

   This immediately leads to the definition of the \emph{Colombeau algebra} as the quotient
\begin{align}
 \label{seq:42}
        \mathcal{G} \DEF \frac{\mathcal{E}_\Mt}{\mathcal{N}}.
\end{align}
That is, an element $g \in \mathcal{G}$ is an equivalence class $[g] = [g_\epsilon + \mathcal{N}]$ of an element $g_\epsilon \in \mathcal{E}_\Mt$, which is called a \emph{representative} of the \emph{generalized function} $g$.  The product $g \odot h$ is defined as the class of $g_\epsilon \cdot h_\epsilon$ where $g_\epsilon$ and $h_\epsilon$ are (arbitrary) representatives of $g$ and $h$; similarly $\rmD g$ is the class of $\rmD g_\epsilon$ if $\rmD$ is any partial differentiation operator.  Therefore, when working in $\mathcal{G}$, all algebraic and differential operations (as well as composition of functions, etc.) are performed component-wise at the level of the representatives $g_\epsilon$.

    $\mathcal{G}$ is an associative and commutative differential algebra because both  $\mathcal{E}_\Mt$ and $\mathcal{N}$ are such.  The two main ingredients which led to its definition are the primacy given to $\mathcal{C}^\infty$ functions, and the use of the Colombeau mollifiers for the embeddings.

\section{Special and general Colombeau algebras}

  Depending on the precise definitions of the moderate and negligible functions, as well as of any further specification constraining the Colombeau mollifiers, there can be many variants of $\mathcal{G}$, even if the domain and range of the generalized functions are simply a subset of $\mathbb{R}$.  There are however two general types of Colombeau algebras: The `special' (or `simple') algebras, and the `general' (also called `full' or `elementary') algebras. 

  For example, let us define a \emph{special Colombeau algebra} of generalized functions on $\Omega \in \mathbb{R}^n$ with value in $\mathbb{C}$.  Then, using the standard multi-index notation
\begin{align}
\label{seq:43}
 \rmD^\alpha  = \frac{\partial^{|\alpha|}}
                   {(\partial x_1)^{\alpha_1} \cdots (\partial x_n)^{\alpha_n}},
\end{align}
where $\alpha \in \mathbb{N}_0^n$ and $|\alpha| = \alpha_1 + \alpha_2 + \cdots \alpha_n$, the distributions will be the partial derivatives $\rmD^\alpha f(\vec{x})$ of the continuous function $f(\vec{x})\in\mathcal{C}(\Omega)$.  A possible definition of $\mathcal{G}^s(\Omega)$, which can easily be adapted to more complicated manifolds, is  as follows:
\begin{definition}[Embedding space] 
\label{seq:defi:1}
Let $\Omega$ be an open set in $\mathbb{R}^n$, let $\epsilon \in ]0,1[$ be a parameter, and let $\eta  \in \mathcal{A}_\infty$ be an arbitrary but fixed Colombeau mollifier. The `embedding space' is the differential algebra 
\begin{align}
\nonumber
    \mathcal{E}^s(\Omega) \DEF  \Bigl\{
     f_{\epsilon} : ~~
     ]0,1[ \times \Omega &\rightarrow \mathbb{C},\\
\label{seq:44}
                             (\epsilon, \vec{x}\,) &\mapsto 
                                        f_{\epsilon}(\vec{x}\,) \Bigr\},
\end{align}
where the sequencies $f_{\epsilon}$ are $\mathcal{C}^\infty$ functions in the variable $\vec{x} \in \Omega$.  The compactly supported distributions\footnote{The embedding of all of $\mathcal{D}'$ is achieved by a more complicated formula based on sheaf theoretic arguments, see \cite[Proposition 1.2.13]{GROSS2001-} .} are embedded in $\mathcal{E}^s$ by convolution with the scaled mollifier $\eta_\epsilon$, i.e.,
\begin{align}
\label{seq:45}
  f_\epsilon(\vec{x}\,) \DEF  \int \frac{dy^n}{\epsilon^n}
            ~\eta\Bigl(\frac{\vec{y}-\vec{x}}{\epsilon}\Bigr) ~f(\vec{y}\,)
               = \int dz^n~\eta(\vec{z}\,) ~f(\vec{x} + \epsilon \vec{z}\,).
\end{align}
\end{definition}
\begin{definition}[Moderate functions] 
\label{seq:defi:2}
The differential subalgebra $\mathcal{E}^s_\Me \subset \mathcal{E}^s$ of `moderate functions' is
\begin{align}
\nonumber
    \mathcal{E}^s_\Me(\Omega) \DEF
 \Bigl\{ f_\epsilon : ~ & \forall K \text{~compact in~} \Omega,
                            \forall \alpha \in \mathbb{N}_0^n,\\
\nonumber
         & \exists N \in \mathbb{N}_0 \text{~such that}, \\
\label{seq:46}
         & \sup_{\vec{x} \in K} ~ |\rmD^\alpha f_\epsilon(\vec{x}\,)|
         = \OOO(\frac{1}{\epsilon^N})
           \text{~~~as~} \epsilon \rightarrow 0
  \Bigr\}.
\end{align}
\end{definition}
\begin{definition}[Negligible functions]
\label{seq:defi:3}
The differential ideal $\mathcal{N}^s \subset \mathcal{E}^s_\Me$ of `negligible functions' is
\begin{align}
\nonumber
    \mathcal{N}^s(\Omega) \DEF
 \Bigl\{ f_\epsilon : ~ & \forall K \text{~compact in~} \Omega,
                            \forall \alpha \in \mathbb{N}_0^n,\\
\nonumber
         & \forall q \in \mathbb{N}, \\
\label{seq:47}
         & \sup_{\vec{x} \in K} ~ |\rmD^\alpha f_\epsilon(\vec{x}\,)|
         = \OOO(\epsilon^q)
           \text{~~~as~} \epsilon \rightarrow 0
  \Bigr\}.
\end{align}
\end{definition}
\begin{definition}[Special algebra] 
\label{seq:defi:4}
The special Colombeau algebra is the quotient 
\begin{align}
\label{seq:48}
        \mathcal{G}^s    (\Omega) \DEF 
 \frac {\mathcal{E}^s_\Me(\Omega)}
       {\mathcal{N}^s    (\Omega)}.
\end{align}
\end{definition}
The main differences with the `naive' definitions \eqref{seq:35} and \eqref{seq:37} are: (i) The resort to the compact subset $K \subset \Omega$, which is necessary because of the local character of the concept of distribution, as is clearly stipulated by Schwartz's structure theorem; (ii) the need to consider the supremum over all $K \subset \Omega$ in order to take into account all possible discontinuities when $x$ ranges in $\Omega$; and (iii) the need to consider all possible derivatives of $f_\epsilon$ in order that the moderate and negligible functions have the required properties for all their derivatives.

    A \emph{general Colombeau algebra} $\mathcal{G}^g$ is an enlargement of $\mathcal{G}^s$, obtained by considering all $\eta \in \mathcal{A}_\infty$ and by replacing (in both $\mathcal{N}^s$ and $\mathcal{E}^s_\Me$) the functions $x \rightarrow f_\epsilon(x)$ by the set of functions $x \rightarrow f_\epsilon(x,\eta)$ depending on $\eta$.  Since all possible $\eta$ are considered the arbitrariness characteristic of $\mathcal{G}^s(\Omega)$ disappears, and the embeddings of the distributions and functions with finite differentiability become `canonical' since they do not depend any more on a fixed mollifier.  However, while this is conceptually interesting from the mathematical point of view, it is not a necessity since the particular mollifier (or set of mollifiers) defining a special Colombeau algebra $\mathcal{G}^s \subset \mathcal{G}^g$ may have a physical interpretation.  For this reason the dependence of the embeddings on the mollifiers is not a defect, but rather a positive feature in many applications of the special Colombeau algebras.

 \section{Interpretation of distributions within $\mathcal{G}$}

   To construct the Colombeau algebra we have been led to embed the distributions as the representative sequences $(\gamma\epsPeps \in \mathcal{E}$ defined by \eqref{seq:17} where $\eta_\epsilon \in \mathcal{A}_\infty$ is a Colombeau mollifier, that is not as the usual representative sequencies defined by \eqref{seq:10} where $\rho_\epsilon \in \mathcal{A}_0$.  However, since $\mathcal{A}_\infty \subset \mathcal{A}_0$, we can still recover any distribution $\gamma$ from $\gamma_\epsilon = (\gamma\epsPeps$ by means of \eqref{seq:11}, i.e., as the equivalence class
\begin{equation}
\label{seq:49}
        \gamma(T) \DEF \lim_{\epsilon \rightarrow 0} 
                 \int dx~\gamma_\epsilon(x) ~T(x),
                \qquad \forall T(x) \in \mathcal{D},
\end{equation}
where $\gamma_\epsilon$ can be any representative of the class $[\gamma] = [\gamma_\epsilon + \mathcal{N}]$ because negligible elements are zero in the limit $\epsilon \rightarrow 0$.

   Of course, as we work in $\mathcal{G}$ and its elements get algebraically combined with other elements, there can be generalized functions $[g_\epsilon]$ different from the class $[\gamma_\epsilon]$ of an embedded distribution which nevertheless correspond to the {same} distribution $\gamma$.  
This leads to the concept of \emph{association}:  We say that two generalized functions $g$ and $h$ are associated, and we write $g \asymp h$,\footnote{In the literature the symbol $\approx$ is generally used for association.  We prefer to use $\asymp$ because association is not some kind of an `approximate' relationship, but rather the precise statement that a generalized function corresponds to a distribution.} iff
\begin{equation}
\label{seq:50}
                 \lim_{\epsilon \rightarrow 0} 
                 \int dx~\Bigl(g_\epsilon(x)-h_\epsilon(x)\Bigr) T(x) = 0,
                 \qquad \forall T(x) \in \mathcal{D}.
\end{equation}
Thus, if $\gamma$ is a distribution and $g$ some generalized function, the relation $g \asymp \gamma$ implies that $g$ admits $\gamma$ as `associated distribution,' and $\gamma$ is called the `distributional shadow' (or `distributional projection') of $g$ because the mapping $\gamma_\epsilon \mapsto \gamma$ defined by \eqref{seq:49} is then a projection $\mathcal{G} \rightarrow \mathcal{D}'$ for all $g_\epsilon$ associated to $\gamma_\epsilon$.

   Objects (functions, numbers, etc.) which are equivalent to zero in $\mathcal{G}$, i.e., equal to $\OOO(\epsilon^q), \forall q \in \mathbb{N}$, are called \emph{zero}.  On the other hand, objects associated to zero in $\mathcal{G}$, that is which tend to zero as $\epsilon \rightarrow 0$, are called \emph{infinitesimals}.  Definition \eqref{seq:50} therefore means that two different generalized functions associated to the same distribution differ by an infinitesimal.

 \section{Multiplication of distributions in $\mathcal{G}$}

  The continuous functions and their derivatives, i.e., the distributions, are not subalgebras of $\mathcal{G}$:  Only the smooth functions have that property.  Thus we do not normally expect that their products in $\mathcal{G}$ will be associated to some continuous functions or distributions:  In general these products will be genuine generalized functions, i.e., new mathematical objects --- which constitute one of the main attractions of $\mathcal{G}$.

   For example, the $n$-th power of Dirac's $\delta$-function in $\mathcal{G}$, Eq.~\eqref{seq:36}, has no associated distribution.  But $\delta^n$ is a moderate function and thus makes perfectly sense in $\mathcal{G}$.  Moreover, its point-value at zero, $\eta^n(0)/\epsilon^n$ can be considered as a `generalized number.'  

   On the other hand, we have elements like the $n$-th power of Heaviside's function, Eq.~\eqref{seq:34}, which has an associated distribution but is such that $[\HU^n](x)\neq[\HU](x)$ in $\mathcal{G}$, whereas $\HU^n(x) = \HU(x)$ as a distribution in $\mathcal{D}'$.  Similarly, we have $[x]\odot[\delta](x)\neq 0$ in $\mathcal{G}$, whereas $x\delta(x)=0$ in $\mathcal{D}'$.  In both cases everything is consistent:  Using \eqref{seq:50}, one easily verifies that indeed $[\HU^n](x) \asymp [\HU](x)$ and $[x]\odot[\delta](x)\asymp 0$. 

  These differences between products in $\mathcal{G}$ and in $\mathcal{D}'$ stem from the fact that distributions embedded and multiplied in $\mathcal{G}$ carry along with them infinitesimal information on their `microscopic structure.'  That information is necessary in order that the products and their derivatives are well defined in $\mathcal{G}$, and is lost when the factors are identified with their distributional projection in $\mathcal{D}'$.

   Let us illustrate this essential point with a concrete example.  In physics the Heaviside function $\HU(x)$ represents a function whose values jump from $0$ to $1$ in a tiny interval of width $\epsilon$ around $x=0$.  Thus it is obvious that $\int \bigl( \HU^2(x) - \HU(x) \bigr) T(x)~dx$ tends to 0 when $\epsilon \rightarrow 0^+$ if $T$ is a bounded function, i.e., $\HU^2 \ASS \HU$.  But since $\HU'$ is unbounded one has $\int \bigl(\HU^2(x) - \HU(x) \bigr).  \HU'(x)~dx = 1/3 - 1/2 = -1/6$, as obvious from elementary calculations.  This shows that one is not allowed to state $\HU^2=\HU$ in a context where the function $\HU^2-\HU$ could be multiplied by a function taking infinite values such as the Dirac function $\delta=\HU'$.

   Therefore, the distinction between $\mathcal{G}$-functions that are `infinitesimally nonzero' such as $\HU^2-\HU$ from the genuine zero function insures that multiplication is coherent in $\mathcal{G}$, because `infinitesimally nonzero quantities,' when multiplied by `infinitely large quantities,' can give significant nonzero results.  At the same time, this distinction insures that all calculations are consistent with those in $\mathcal{D}'$.  In particular, if at any point it is desirable to look at the intermediate results of a calculation from the point of view of distribution theory, one can always use the concept of association to retrieve their distributional content.  In fact, this is facilitated by a few simple formulas which easily derive from the definition \eqref{seq:50}.  For instance,
\begin{align}
\label{seq:51}
   \forall f_1,\forall f_2 \in \mathcal{C}
   \qquad &\Rightarrow \qquad
   [f_1]\odot[f_2] \asymp [f_1\cdot f_2],\\
\intertext{and,}
\label{seq:52}
   \forall f \in \mathcal{C}^\infty,\forall \gamma \in \mathcal{D}'
   \qquad &\Rightarrow \qquad 
   [f]\odot[\gamma] \asymp [f\cdot \gamma],\\
\intertext{but, in general,}
\label{seq:53}
   \forall \gamma_1,\forall \gamma_2 \in \mathcal{D}'
   \qquad &\Rightarrow \qquad 
   [\gamma_1]\odot[\gamma_2] \not\asymp [\gamma_1\cdot \gamma_2],\\
\intertext{whereas,}
\label{seq:54}
      \forall g_1,\forall g_2 \in \mathcal{G}, \qquad
      g_1 \asymp g_2 \qquad &\Rightarrow
                     \qquad \rmD^\alpha g_1 \asymp \rmD^\alpha g_2.
\end{align}
For example, applying the last equation to $[\HU^2](x) \asymp [\HU](x)$ one proves the often used distributional identity $2[\delta](x)[\HU](x) \asymp [\delta](x)$.

   In summary, one calculates in $\mathcal{G}$ as in $\mathcal{C}^\infty$ by operating on the representatives $g_\epsilon \in \mathcal{E}$ with the usual operations $\{ +,-,\times,d/dx\}$.  The distributional aspects, if required, can be retrieved at all stages by means of association.

\section{Working with distributions versus working with $\mathcal{G}$-functions}
\label{wor:0}

   An interesting feature of Colombeau algebras is that they enable, in many cases,  to set aside the concept of distributions and to replace it by the more general and flexible one of $\mathcal{G}$-functions.

   Indeed, a distribution cannot be the end result of a calculation in any physical theory:  It is a functional which has to be integrated over its argument to yield a quantity comparable to experiment.  Similarly, a measurement is always some kind of an average over a continuous distribution of matter supported by bodies of finite extension.  Thus, if one takes the sequential view, one is often led in physics to consider integrals of the type \eqref{seq:11}, i.e.,
\begin{equation}
\label{wor:1}
        g(S) = \lim_{\epsilon \rightarrow 0} 
                 \int dx~g_\epsilon(x) ~S(x),
\end{equation}
where $g(x)$ may be any regular or singular distribution corresponding to a basic physical quantity (e.g., an energy density), and $S(x) \in \mathcal{D}$ a smooth function (e.g., $S(x)dx$ could be a volume element).

  There are then two options:
\begin{itemize}
\item In conventional `distribution theory' the distributional aspect is emphasized throughout the calculation and all intermediate results are interpreted as distributions.  In particular, when working according to the sequential view, limits similar to that in \eqref{wor:1} are taken at all stages so that information that could be relevant to nonlinear operations is discarded. (In the language of generalized functions, one systematically works with the distributional shadows rather than with the generalized functions themselves.)  The method is therefore restricted to linear theories, and if the limit $\epsilon \rightarrow 0$ is undefined the end result will in general be meaningless even if $\epsilon$ is kept finite. (Because infinitesimal information that could have been significant before passing to the limit may have been discarded.)

\emph{Example:}  If the electrostatic Coulomb potential is defined as a distribution, the $\mathcal{G}$-embedded Coulomb field has the form \cite{TEMPL1953-,GSPON2006B2, GSPON2008B}
\begin{align}
\label{wor:2}
  \vec{E}_\epsilon(\vec{r}\,)
       &= e \lim_{a \rightarrow 0}
            \bigl(\frac{\vec{r}}{r^3} \HU(r-a)\bigr)_\epsilon
        - e \lim_{a \rightarrow 0}
            \bigl(\frac{\vec{r}}{r^2} \DUP(r-a)\bigr)_\epsilon \\
\label{wor:3}
    &\ASS e \lim_{a \rightarrow 0} \frac{\vec{r}}{r^3} \HU(r-a).
\end{align}
In distribution theory, the second part of \eqref{wor:2}, which contains the $\delta$-function, is discarded \cite[p.\,144]{TEMPL1953-}.  Expression \eqref{wor:3} cannot therefore be used in non-linear calculations such as the self-energy of an electron.

\item In `$\mathcal{G}$-function theory' all non-smooth functions $f$ are represented by their Colombeau mollified sequence $f_\epsilon$, and there is a unique parameter $\epsilon$ which is kept finite until the end of the calculation.  (There is also possibly a unique common mollifier $\eta$ if one works in a special Colombeau algebra.)  All operations are then performed on these representatives, and at any stage one can verify the validity of the calculations by checking that the intermediate results are moderate functions. It is possible to consistently manipulate singular distributions in nonlinear calculations, and the end results are obtained by taking the limit $\epsilon \rightarrow 0$ as in \eqref{wor:1}.  If the theory is linear, these results are identical to those of the conventional `distribution theory' option.  In linear or nonlinear theories which lead to divergent quantities as $\epsilon \rightarrow 0$ the parameter $\epsilon$ can be left finite, and the end result can be interpreted as a `generalized number.'  This generalized number may then be renormalized to some finite quantity, which implies that any dependence on $\epsilon$ and on the arbitrary mollifier $\eta$ is removed at this final stage.

\emph{Example:}  Calculating the self-energy of an electron involves integrating the square of its electric field.  In distribution theory, where this square is undefined, the result using \eqref{wor:3} is the well-known expression
\begin{equation} \label{wor:4}
                U_{\text{self}}(\mathcal{D}')
                 = \frac{e^2}{2}\lim_{a \rightarrow 0} \frac{1}{a},
\end{equation}
which diverges as $a \rightarrow 0$, and which corresponds to the square of the \emph{first} term in  \eqref{wor:2}, i.e., to the energy of the field surrounding the electron.  On the other hand, when calculated in $\mathcal{G}$ using \eqref{wor:2} the self-energy is  \cite{GSPON2006B2, GSPON2008B}
\begin{align}
\label{wor:5}
 U_{\text{self}}(\mathcal{G})  = \frac{e^2}{2} \lim_{\epsilon \rightarrow 0}
                    \frac{1}{\epsilon} \int_{-\infty}^{+\infty} dx~ \eta^2(-x).
\end{align}
which is independent of the `cut-off' $a$, and which corresponds to the square of the \emph{second} term in \eqref{wor:2}, i.e., to the square of a $\delta$-function.  Thus, when calculated in $\mathcal{G}$, the self-energy is entirely located at the position of the electron, i.e, precisely where its `inertia' as a point-mass is supposed to reside.  It remains therefore to renormalize $U_{\text{self}}(\mathcal{G})$ to the measured mass of the electron, and everything makes mathematically and physically sense. 

\end{itemize}

   That discussion permits to conclude this paper by an analogy:  The relations of the usual functions and distributions to the $\mathcal{G}$-functions are somewhat analogous to those of the real to the complex numbers.  If one looks at  $\epsilon$ as an analog of $i$, then taking the real part of a complex number corresponds to restricting a generalized function to its associated function or distribution by taking the limit $\epsilon \rightarrow 0$.  Working in $\mathbb{R}$ or $\mathcal{D}'$ is therefore less general and flexible than working in $\mathbb{C}$ or $\mathcal{G}$.


\section*{Acknowledgments}

The author is very much indebted to Drs.\ G.\ Falquet and J.-P.\ Hurni for stimulating discussions and comments and for a critical reading of the manuscript.

\newpage

\section*{Appendix: Special algebra of tempered $\mathcal{G}$-functions}

\setcounter{section} {18}                       
\setcounter{equation}{ 0}                       

In many mathematical and physical applications the restriction of the test functions to the space $\mathcal{D}$ of $\mathcal{C}^\infty$ functions with compact support is too restrictive  \cite{SCHME1990A}.  This is the case of the Fourier transform which, in its simplest form, has as a kernel $\cos(px)$ which is not integrable over the whole space $\Omega=\mathbb{R}$.  Thus, just like in Schwartz distribution theory, an extension of the Colombeau theory to `tempered $\mathcal{G}$-functions,' see, e.g., \cite[p.\,15 and p.\,65]{GROSS2001-}, is essential when dealing with functions that are integrated over the whole space $\Omega=\mathbb{R}^n$.  Moreover, the algebra $\mathcal{G}^t$ of tempered $\mathcal{G}$-functions has a property that is important from a practical point of view: In $\mathcal{G}^t$, componentwise composition is a well defined operation generalizing composition of $\mathcal{C}^\infty$ functions.

To use this extension  it is necessary to be careful about the definitions of the pertinent function spaces.  We therefore recall \cite[p.\,15]{GROSS2001-}:  
\begin{definition}[Algebras $\mathcal{S}$, $\mathcal{O}_{\text{C}}$, and $\mathcal{O}_{\text{M}}$ ] 
\label{seq:defi:5}
Let $\Omega^t \subset \mathbb{R}^n$ be a $n$-dimensional box, i.e., a subset of the form $\omega_1 \times \cdots \times \omega_n$ where $\omega_i$ is a finite or infinite open interval in $\mathbb{R}$. Then,
\begin{align}
\nonumber
    \mathcal{S} \DEF  \Bigl\{
     f \in \mathcal{C}^\infty(\Omega^t) : ~
      &\forall m \in \mathbb{N}, \forall \alpha \in \mathbb{N}_0^n,\\
\label{seq:56}
      &\sup_{\vec{x} \in \Omega^t} ~ (1 + |\vec{x}\,|)^{+m}
                                |\rmD^\alpha f(\vec{x}\,)|  < \infty
                      \Bigr\},
\end{align}
\begin{align}
\nonumber
    \mathcal{O}_{\text{\emph{C}}} \DEF  \Bigl\{
     f \in \mathcal{C}^\infty(\Omega^t) : ~
      &\exists m \in \mathbb{N} \text{~such that,~}
       \forall \alpha \in \mathbb{N}_0^n,\\
\label{seq:57}
      &\sup_{\vec{x} \in \Omega^t} ~ (1 + |\vec{x}\,|)^{-m}
                                |\rmD^\alpha f(\vec{x}\,)|  < \infty
                      \Bigr\},\\
\nonumber
    \mathcal{O}_{\text{\emph{M}}} \DEF  \Bigl\{
     f \in \mathcal{C}^\infty(\Omega^t) : ~
      &\forall \alpha \in \mathbb{N}_0^n, \exists m \in \mathbb{N},\\
\label{seq:58}
      &\sup_{\vec{x} \in \Omega^t} ~ (1 + |\vec{x}\,|)^{-m}
                                |\rmD^\alpha f(\vec{x}\,)|  < \infty
                      \Bigr\}.
\end{align}

\end{definition}
$\mathcal{O}_{\text{C}}$ and $\mathcal{O}_{\text{M}}$ correspond to two closely related definitions of functions with polynomial growth as $|\vec{x}\,| \rightarrow \infty$.  But, while $\mathcal{O}_{\text{M}}$ corresponds to the usual definition, it is the algebra $\mathcal{O}_{\text{C}}$ which in the $\mathcal{G}$-context provides the proper `tempered' extension of the notion of $\mathcal{C}^\infty$ functions with compact support.
\begin{definition}[Embedding space of temperate distributions] 
\label{seq:defi:6}
Let $\Omega^t \subset \mathbb{R}^n$ be a $n$-dimensional box, let $\epsilon \in ]0,1[$ be a parameter, and let $\eta  \in \mathcal{A}_\infty \subset \mathcal{S}$ be an arbitrary but fixed Colombeau mollifier.  The `embedding space' is the differential algebra 
\begin{align}
\nonumber
    \mathcal{E}^t(\Omega^t) \DEF  \Bigl\{
     f_{\epsilon} : ~~
     ]0,1[ \times \Omega^t &\rightarrow \mathbb{C},\\
\label{seq:59}
                             (\epsilon, \vec{x}\,) &\mapsto 
                                        f_{\epsilon}(\vec{x}\,) \Bigr\},
\end{align}
where the sequencies $f_{\epsilon}$ are $\mathcal{C}^\infty$ functions in the variable $\vec{x} \in \Omega^t$.  The tempered distributions $g \in \mathcal{S}'$ are embedded in $\mathcal{E}^t$ by convolution with the scaled mollifier $\eta_\epsilon$, i.e.,
\begin{align}
\label{seq:60}
  g_\epsilon(\vec{x}\,) \DEF  \int \frac{dy^n}{\epsilon^n}
            ~\eta\Bigl(\frac{\vec{y}-\vec{x}}{\epsilon}\Bigr) ~g(\vec{y}\,)
               = \int dz^n~\eta(\vec{z}\,) ~g(\vec{x} + \epsilon \vec{z}\,).
\end{align}
The functions $h \in \mathcal{O}_{\text{\emph{C}}}$ are directly embedded, i.e.,
\begin{align}
\label{seq:61}
   \forall h \in \mathcal{O}_{\text{\emph{C}}}, \qquad
   (h\epsPeps(x)  =  h(x),
\end{align}
so that $\mathcal{O}_{\text{\emph{C}}}$ is a subalgebra of $\mathcal{E}^t(\Omega^t)$.

\end{definition}
\begin{definition}[Temperate moderate functions] 
\label{seq:defi:7}
The differential subalgebra $\mathcal{E}^t_\Me \subset \mathcal{E}^t$ of `moderate functions' is
\begin{align}
\nonumber
    \mathcal{E}^t_\Me(\Omega^t) \DEF
 \Bigl\{ f_\epsilon : ~ & \forall K \text{~compact in~} \Omega^t,
                            \forall \alpha \in \mathbb{N}_0^n,\\
\nonumber
         & \exists N \in \mathbb{N}_0 \text{~such that}, \\
\label{seq:62}
         & \sup_{\vec{x} \in K} ~ (1 + |\vec{x}\,|)^{-N}
                                |\rmD^\alpha f_\epsilon(\vec{x}\,)|
         = \OOO(\frac{1}{\epsilon^N})
           \text{~~~as~} \epsilon \rightarrow 0
  \Bigr\}.
\end{align}
\end{definition}
\begin{definition}[Temperate negligible functions]
\label{seq:defi:8}
The differential ideal $\mathcal{N}^t \subset \mathcal{E}^t_\Me$ of `negligible functions' is
\begin{align}
\nonumber
    \mathcal{N}^t(\Omega^t) \DEF
 \Bigl\{ f_\epsilon : ~ & \forall K \text{~compact in~} \Omega^t,
                            \forall \alpha \in \mathbb{N}_0^n,\\
\nonumber
         & \exists m \in \mathbb{N} \text{~such that,~}
           \forall q \in \mathbb{N}, \\
\label{seq:63}
         & \sup_{\vec{x} \in K} ~ (1 + |\vec{x}\,|)^{-m}
                                |\rmD^\alpha f_\epsilon(\vec{x}\,)|
         = \OOO(\epsilon^q)
           \text{~~~as~} \epsilon \rightarrow 0
  \Bigr\}.
\end{align}
\end{definition}
\begin{definition}[Special algebra of tempered $\mathcal{G}$-functions] 
\label{seq:defi:9}
The special Colombeau algebra is the quotient 
\begin{align}
\label{seq:64}
        \mathcal{G}^t    (\Omega^t) \DEF 
 \frac {\mathcal{E}^t_\Me(\Omega^t)}
       {\mathcal{N}^t    (\Omega^t)}.
\end{align}
\end{definition}
%



\end{document}